\documentclass{article}

\usepackage{arxiv}

\usepackage[utf8]{inputenc} 
\usepackage[T1]{fontenc}    
\usepackage{hyperref}       
\usepackage{url}            
\usepackage{booktabs}       
\usepackage{amsfonts}       
\usepackage{nicefrac}       
\usepackage{microtype}      
\usepackage{lipsum}

\usepackage{graphicx}
\usepackage{subfigure}
\usepackage{multirow}

\title{Using infrared gas sensors in an in-vitro dynamic gut model for detecting short-chain fatty-acids: Technical Report }

\author{
  Martin Länkvist \\
  Center for Applied Autonomous Sensor Systems\\
  \"Orebro University\\
  \"Orebro, Sweden \\
  \texttt{martin.langkvist@oru.se}
   \And
  Amy Loutfi \\
  Center for Applied Autonomous Sensor Systems\\
  \"Orebro University\\
  \"Orebro, Sweden \\
  \texttt{amy.loutfi@oru.se} 
     \And
  Ignacio Rangel \\
  Örebro Life Science Center\\
  \"Orebro University\\
  \"Orebro, Sweden \\
  \texttt{ ignacio.rangel@oru.se}
     \And
  Johnny Karlsson \\
  Örebro Life Science Center\\
  \"Orebro University\\
  \"Orebro, Sweden \\
  \texttt{johnny.karlsson@oru.se}
     \And
  Robert Jan Brummer \\
  Örebro Life Science Center\\
  \"Orebro University\\
  \"Orebro, Sweden \\
  \texttt{robert.brummer@oru.se}
}

\begin{document}
\maketitle

\begin{abstract}

\end{abstract}


\section{Introduction}

Short-chain fatty acids (SCFAs), including acetate, propionate and butyrate, are organic fatty-acids that are produced when indigested carbohydrates are fermented in the colon by gut-bacteria. Butyrate is especially considered a beneficial compound in relation to gut health and the maintenance of colonic homeostasis~\cite{hamer2008role}. Little is known about butyrate production and the current measurement methods of fecal samples are not representative and there is a strong unmet need to measure SCFAs \emph{in vivo} to further understand their role and effect on colonic function in human gut health and disease. 

The general aim of the project is to develop a novel sensor that is capable of online detecting short chain fatty acids (SCFA; e.g. acetate, propionate and butyrate) and other small biomolecules (e.g. ammonium) produced in bioreactors.
Whereas currently such levels of small biomolecules are usually not detected online due to practical constraints (sampling-analysis-processing of data), reliable online detection could significantly decrease analysis costs. Further, it would allow to perform online quality control of the bioreactors and allow to e.g. detect deviating values at an early stage. Hence, one can react much faster tackling certain problems which will further improve the quality of each experiment.

\section{Materials and Methods}
\label{sec:method}

\subsection{Infrared gas sensors}

There are multiple advantages to using an Infrared (IR) sensor for this application instead of the commonly academically used MOS sensors. The first advantage is that MOS sensors have limited selectivity and require multiple differently doped MOS sensors combined with pattern recognition algorithms to identify the target gas, while an IR sensor has a specific selectivity towards a target gas set by the frequency of the LED lamp. Figure 10 shows the functionality of an IR sensor. It consists of an IR lamp set to a certain frequency and an IR detector. The target gas is passed between the lamp and the detector and certain molecules will absorb the infrared light and the detector will detect the difference between the actual received energy and the expected received energy. The optical path, L, determines the sensitivity (long path = low gas concentrations and vice versa). Figure 11 shows the amount of absorption for different wavelength of the LED lamp. Many molecules, such as NO, CO, and CO2 absorb the infrared light at various wavelength. One of the advantages of IR technology is that these sensors will only react to the target gas independent of what other gases are present. But the requirement is that the target molecule must be able to have multiple di-pole states, which makes gases such as H2, N2, and O2 undetectable. To our knowledge, it is unknown if the target SCFA gases for our application absorb IR light or not, but if it did it would allow for maximum selectivity since it would eliminate the influence of other present gases. IR technology has the advantages of being used in aggressive chemical environments, anaerobic conditions, less calibration and burn-out, long life expectancy, and no cross-sensitivity. The disadvantages are that it is very important that water does not condense or leak into the sensor and the bulkiness of the fragile system of both a lamp and a detector. 

IR technology is an optical remote sensing technology that doesn’t require the gas to be in physical contact with the target gases. This opens up the possibility for ex vivo remote sensing. Before investigating if this approach is feasible, the experiments with the SHIME system need to determine if IR sensors are capable of detection our target gases, or alternatively, gases that are correlated to our target gases.

Best wavelength for acetate is argmax(A-B-P) = 624.6
Best wavelength for buterate is argmax(B-A-P) = 1231.2
Best wavelength for propionate is argmax(P-A-B) = 1129.1

\begin{figure}
  \centering
  \includegraphics[width=0.7\textwidth]{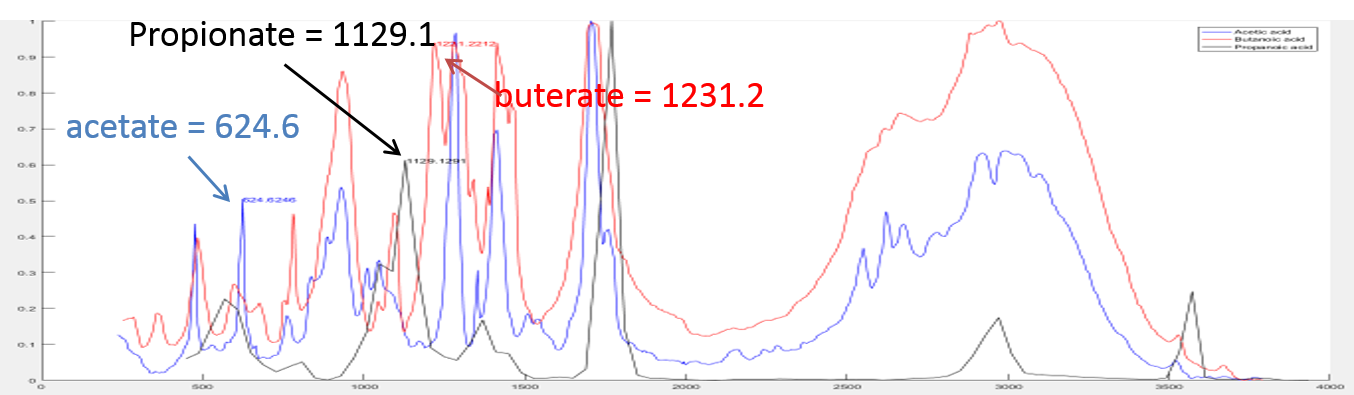}
  \caption{Infrared spectrums for butyrate, acetate, and propionate from the NIST Chemistry WebBook at \url{http://webbook.nist.gov/chemistry/}. }
  \label{fig:irsensordata}
\end{figure}

Infra-red (IR) sensor (gives 2 signals)

\begin{table}[!h]
\centering
\begin{tabular}{c | c | c}
Sensor & Responsive gas & Detection range \\ \hline
IR15TT & $CO_2$ & $0 - 2\%$ \\
       & $CH_4$ & $0 - 5\%$ \\
       & Hydrocarbons &  \\
   \end{tabular}
   \caption{Details about the type of sensors used in the electronic nose sensor array that was used in this work and their target gas and the detection range.}
   \label{table:sensorarray}
\end{table}

\begin{figure}
  \centering
  \includegraphics[width=0.7\textwidth]{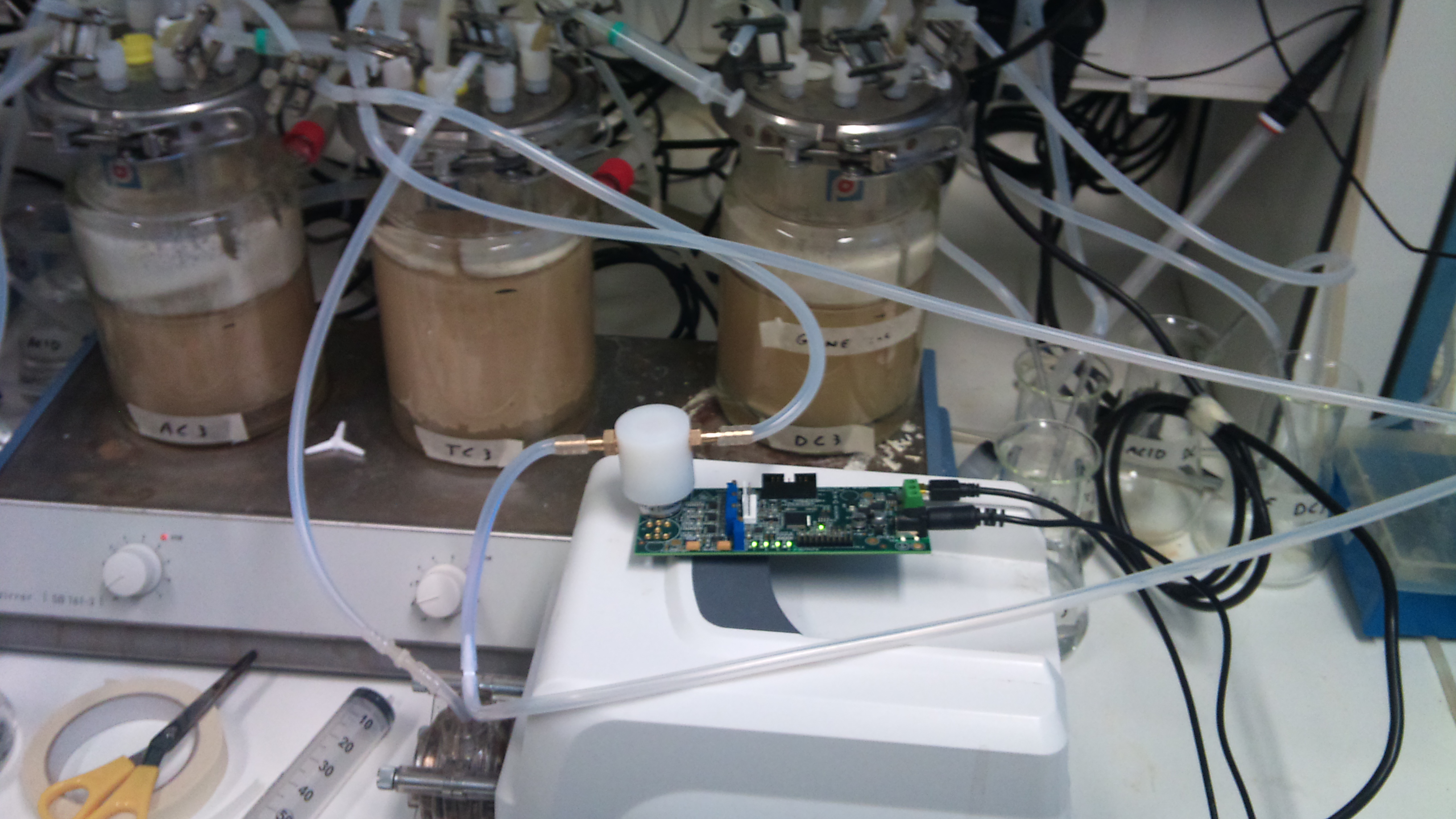}
  \caption{Our sensor setup connected to the SHIME system.}
  \label{fig:shimeirsensor}
\end{figure}

\subsection{Simulator of the Human Intestinal Microbial Ecosystem (SHIME)}
\label{sec:shime}

\begin{figure}
  \centering
  \includegraphics[width=0.7\textwidth]{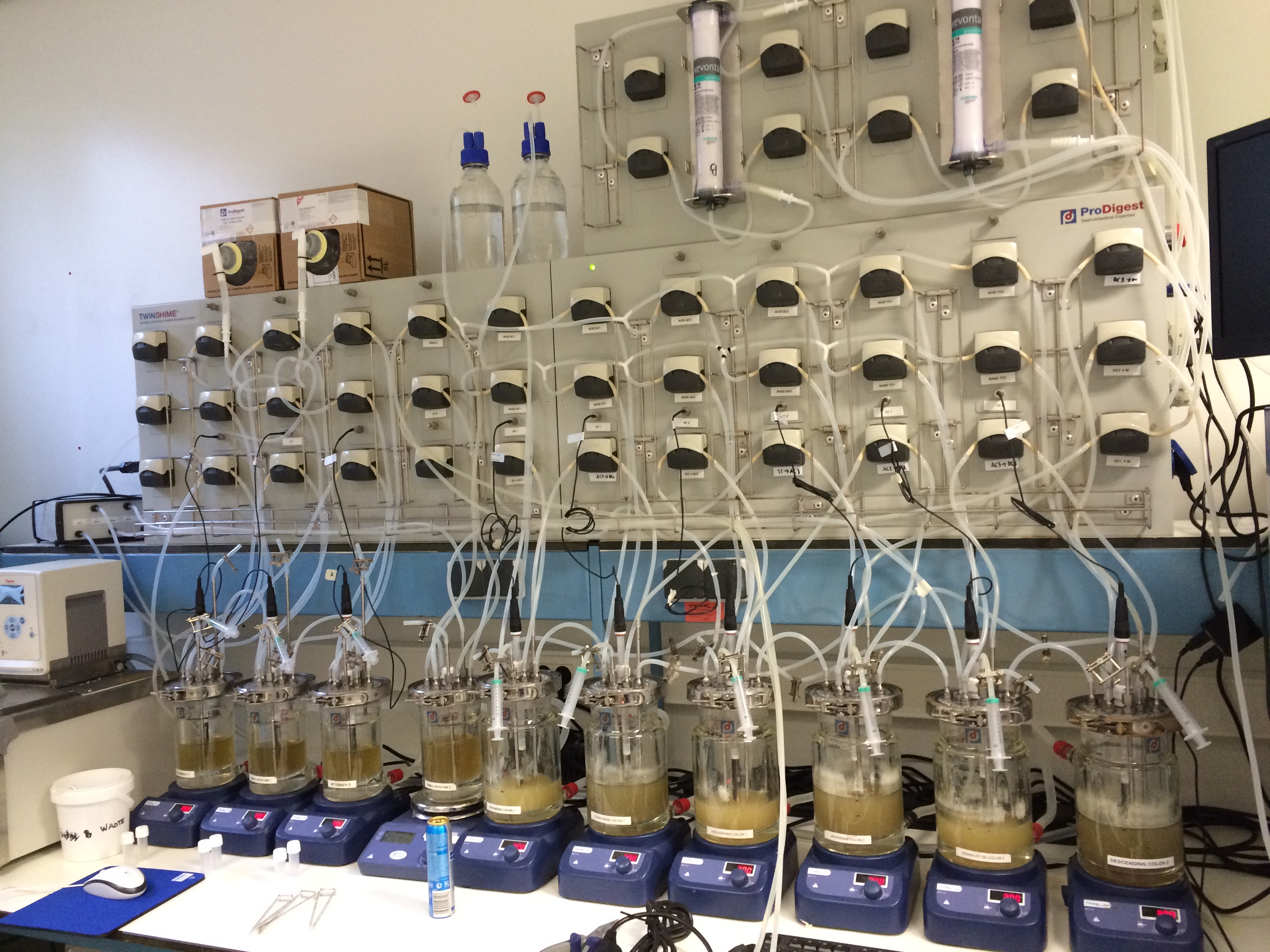}
  \caption{SHIME gut model at ProDigest, Ghent, Belgium.}
  \label{fig:shime}
\end{figure}

ProDigest is a spin-off company from the Laboratory of Microbial Ecology and Technology (LabMET), located at Ghent University (Belgium).

ProDigest's services in the field of gastrointestinal transit, bioavailability and metabolism are based on the application of an extensive in vitro technology platform which was developed at LabMET and which is organized round a dynamic simulation model of the gastrointestinal tract, called the Simulator of the Human Intestinal Microbial Ecosystem (SHIME). This technology platform is combined with a wide range of in vitro assays, animal trials and proof-of-concept human intervention studies.

\subsection{Experimental setup}

Data was collected at ProDigest in the beginning of spring 2016. A MOS sensor array and IR sensor were tested in the SHIME system. Short-term experiments revealed that the anaerobic conditions of the system saturated the reading of the MOS sensors. The IR sensor was therefore used for a long-term experiment. The setup can be seen in Figure 10. A dual-channel IR sensor (uniquely sensitive to CO2 and hydrocarbons) was connected with a closed-loop to two vessels that are fed different nutritients three times a day. The ground truth SCFA-level of acetate, propionate, and butyrate is obtained with a GC-MS before each feeding. The vessels are flushed with nitrogen gas once a day to remove oxygen and the sensor is flushed before each reading. The IR sensor reads continuosly each vessel for 55 minutes before and after each feeding. The experiment is carried out for a period of 96 hours.

The reactor setup was adapted from the SHIME, representing the gastrointestinal tract (GIT) of the adult human, as described by~\cite{molly1993development}. The SHIME consists of a succession of five reactors simulating the different parts of the human gastrointestinal tract (Fig. 1). The first 2 reactors simulate the stomach and the small intestine. In the stomach, a defined amount of SHIME nutritional medium, simulating food intake (as described more in detail below), is added. After this, a standardized pancreatic and bile liquid is added for the simulation of small intestinal conditions. Incubation conditions, retention time and pH are chosen in order to resemble in vivo conditions in the upper part of the gastrointestinal tract.

To test a first prototype of the biosensor two parallel proximal colon (PC) reactors were compared with each other. Whereas one SHIME received the conventional nutritional medium (PC1; control), the second will receive a butyrogenic fiber treatment used during the fibebiotics project (PC2; NAXUS).

The first unit (control) was provided with the conventional SHIME nutritional medium [Arabic gum (1.2 g/L), Pectin (2 g/L), Xylan (0.5 g/L), Starch (4 g/L), Glucose (0.4 g/L), Yeast extract (3 g/L), Peptone (1 g/L), Mucin (3 g/L), Cysteine (0.5 g/L)]. To increase butyrate production, the second SHIME unit was treated with 10 g/L NAXUS in addition to the SHIME nutritional medium. Upon addition of the nutritional medium, gastric suspensions were incubated at pH 2 during 1h30min.
Small intestinal pH values are considerably higher than gastric pH values due to the neutralization of the gastric suspension with bicarbonate ions secreted by the pancreas. In this experiment a standardized pancreatic/bile juice was used [NaHCO3 (12.5 g/L), Bile (6 g/L), Pancreatin (0.9 g/L)]. Residence time in small intestine was 1h30min.
Three times per day small intestinal fluids entered the proximal colon (every 8 hours). The pH of the proximal colon was maintained stable between 5.6 and 5.9. More detailed timing information of this experiment can be found in figure 2.

Experiments:
1.IR sensor in closed-loop with one vessel (long and short)
2.IR and MOS sensor in closed-loop with one vessel (short)
3.IR sensor in closed-loop with two vessels (long ongoing)

\section{Experimental Results}
\label{sec:results}

Consists of a pump that creates a circular airflow between the headspace and the sensors. The system is closed so no gas can come in or out. A flushing system is connected to flush N2 gas (99.9\%) through the sensors. A separate flushing loop is connected to only flush the headspeace of the vessel (not shown in picture).

\begin{figure}
  \centering
  \includegraphics[width=0.7\textwidth]{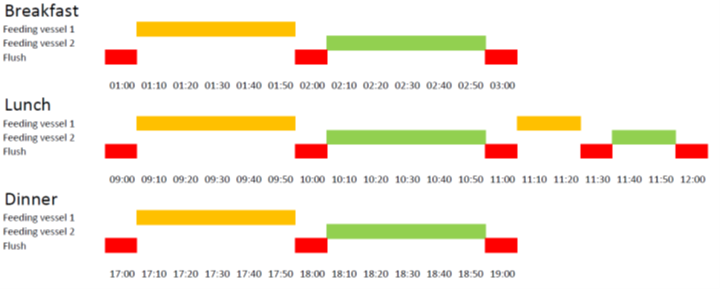}
  \caption{Gas response from IR and MOS sensors, with gas flow. During flush action 1, 2, 3 and 4 all SHIME vessels are flushed. Flush action 5 flushes the biosensor. Different pump actions transfer either feed, pancreatic juice (PJ) or colon suspension to the appropriate vessels}
  \label{fig:shimeschedule}
\end{figure}


IR and MOS sensor in serial in closed-loop with one vessel. Experiment consists of two flushings and two pumpings. IR sensors and humidity sensors behave as expected. But two of the MOS sensors seem to be faulty and the other two get saturated when the pumping starts.

Problem: The holes for the caps for MOS sensors were too small. The pressure from the SHIME pumps where not strong enough to pump any gas through.
Attempted solution: A plastic bag with a sealed input and output that covers all the sensors were tried instead of the caps.
Result: Even with great effort the bag leaked gas and the sensors seem damaged or saturated when exposed to the vessel headspace. A new solution for MOS sensors is needed.

\begin{figure}
  \centering
  \includegraphics[width=0.7\textwidth]{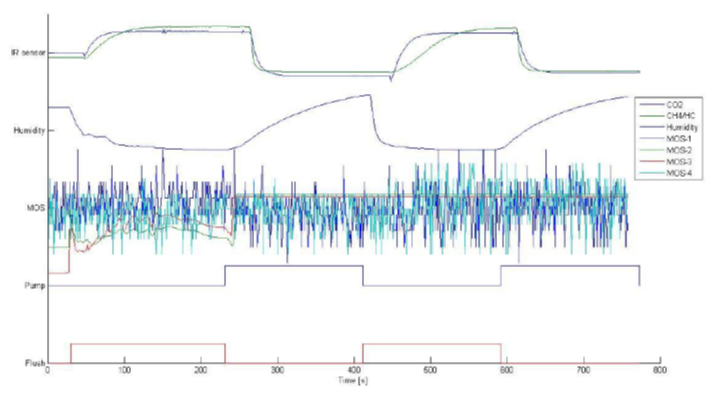}
  \caption{Gas response from IR and MOS sensors, with gas flow.}
  \label{fig:shimedata}
\end{figure}

IR sensors are connected to the SHIME system in closed-loop with two vessels. 

To avoid saturation, the biosensor was only exposed to the gas phase of each colon vessel during 55 min rather than continuously exposing the biosensor to this gas. Subsequently, in order to register a baseline before each reading, the sensor was flushed with nitrogen gas during 5min:

Both vessels are ”feed” 3 times per day with different diets. After some weeks introduce anti-biotika. Only IR sensor is used.

\begin{figure}
  \centering
  \includegraphics[width=0.7\textwidth]{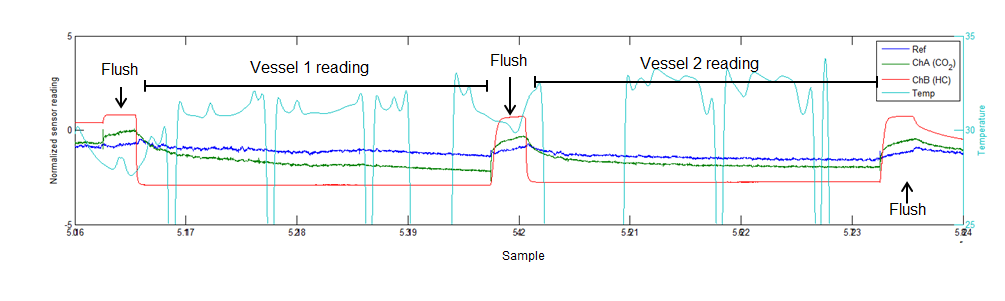}
  \caption{A short window of sensor readings of both vessels over a duration of 2 hours. }
  \label{fig:shimedata2vessel}
\end{figure}

A 2-hour long reading from the IR-sensor is shown in Figure~\ref{fig:shimedata2vessel}. The sensor provides 4 signals: the $CO_22$ level, hydro-carbon (HC) level, temperature, and a reference signal. The temperature reading is not used in the analysis. It can be seen that the flushing operation restores the sensor readings to their baseline values and that the signals reach different final values for the two different vessels.

\begin{figure}
  \centering
  \includegraphics[width=0.7\textwidth]{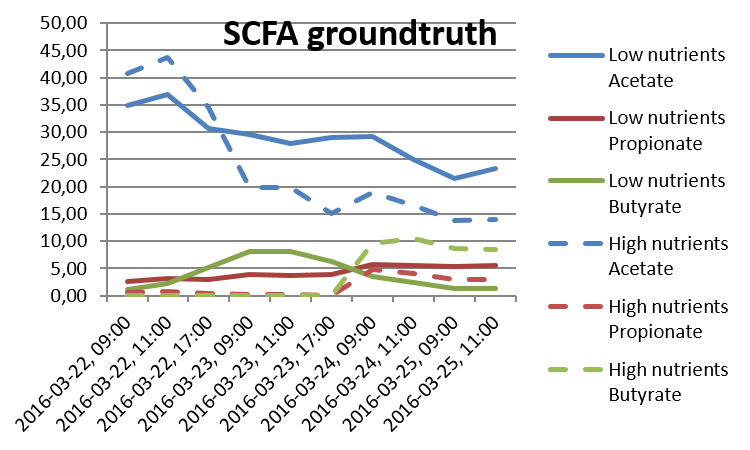}
  \caption{Groundtruth of SCFA-levels for a long-time experiment.}
  \label{fig:shimegroundtruth}
\end{figure}

The groundtruth for the SCFA-levels can be seen in Figure~\ref{fig:shimegroundtruth}.

The pH-level can be seen in Figure~\ref{fig:shimegroundtruth}.
\begin{figure}
  \centering
  \includegraphics[width=0.7\textwidth]{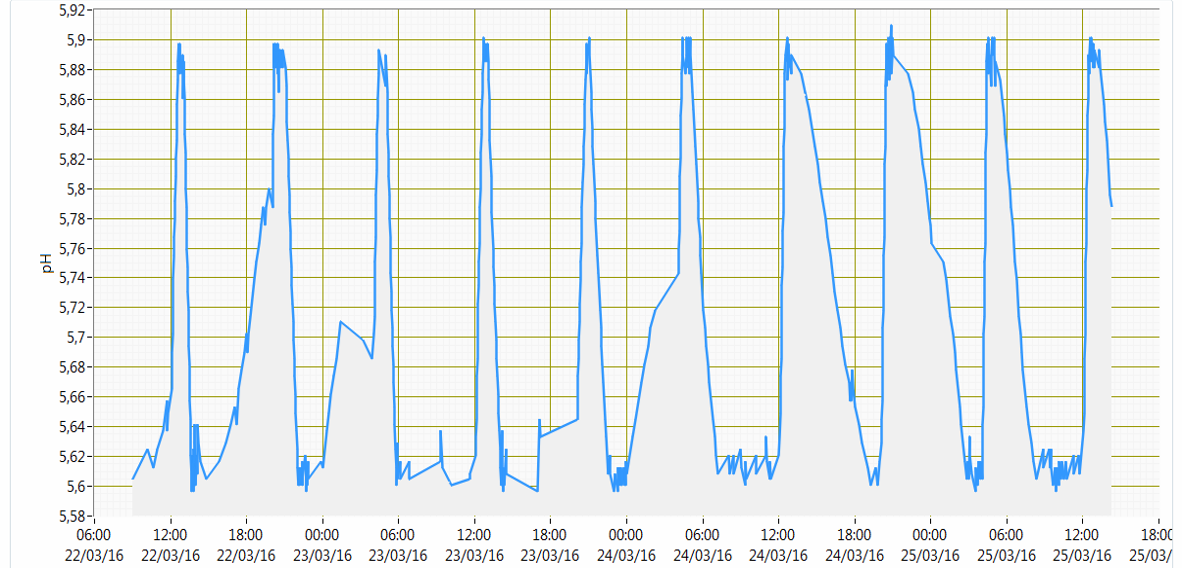}
  \caption{pH level for a long-time experiment.}
  \label{fig:shimeph}
\end{figure}

A machine learning algorithm was implemented to test if the signals could be used to predict the level of each SCFA. The level for each substance was categorized into two classes: high level and low level. A high level is defined as above the average concentration over the 4-day period and vice versa. A total of 12 features was extracted from each signal (ref, ChA, and ChB) in each reading, giving a total of 36 features. A feature selection algorithm was used to select the best features for each substance. The complete data set contained 19 readings with known ground truth. A full leave-one-out cross-validation was performed and the result is shown in Table~\ref{table:shimeresult}. The number of correctly classified high or low SCFA-level was 17, 14, and 10 readings for acetate, propionate, and butyrate, respectively. The easiest substance to quantify was acetate, followed by propionate and the hardest was butyrate with just over half of the readings correctly classified. The best features for predicting the level of acetate mostly contained features from the CO2 signal while the best features for predicting butyrate came from the HC signal. This shows that different signals are good at predicting different substances. The analysis can be extended to include more levels to be classified such as low, medium, and high or treat the problem as a regression problem to predict the value of the concentration. The experiement will be repeated for an even longer duration and with additional IR-sensors in order to ascertain if IR-sensor technology can be used to reliably predict butyrate in particular.

\begin{figure}
  \centering
  \includegraphics[width=0.7\textwidth]{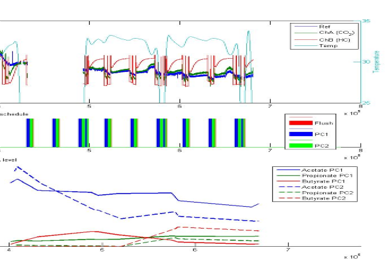}
  \caption{Gas response from IR sensors for a long-time experiment.}
  \label{fig:shimedatalong}
\end{figure}

\begin{table}[!h]
\centering
\begin{tabular}{c | c | c}
Substance	& Accuracy on predicting the concentration & Top features\\ \hline
Acetate	& 89.5 (17/19) & $ref_{std}$, $ChA_{ampdiff}$, $ChA_{std}$, $ChA_{dwtStd}$ \\
Propionate	& 73.7 (14/19) & $ChA_{dwtStd}$, $ChA_{std}$, $ref_{ampdiff}$, $ChB_{ampdiff}$ \\
Butyrate	& 52.6 (10/19) & $ref_{ampdiff}$, $ChB_{kurt}$, $ChB_{median}$, $ref_{kurt}$ \\
   \end{tabular}
   \caption{Accuracy on predicting the correct substance and concentration (high or low SCFA-level) during a long-time experiment.}
   \label{table:shimeresult}
\end{table}

During start-up a 500ml of nutritional medium was inoculated with a human fecal sample. To allow the bacteria to grow to a dense culture, there was no addition of fresh nutritional medium during the overnight incubation, thus resulting in a strong SCFA production. At 9h, the concentration was already 40 mmol/L in the control unit (PC1) while still increasing until 11h (Figure 4). At 12h, the first feeding from the small intestine entered the colon. After 13h all the small intestinal fluids had entered the colon diluting the present SCFA concentration (figure 4A). Because the bacteria are not able to instantaneously ferment the new fibers upon each feeding there is a delay between adding new medium and SCFA production. SCFA concentration will thus always have small fluctuations over time.
Unfortunately there were some technical problems with pH regulation of the second SHIME. With the entire experiment being set up on one day, a small mistake was made while writing the program. To set up the pH controllers of PC1 and PC2, the same pH probe (the one of PC1) was assigned in both cases. As a result, pH control for PC1 was correct, however, pH control of unit 2 was done based on measurements in PC1. As the second unit was treated with NAXUS (thus encountering stronger microbial fermentation and SCFA production), base consumption was not sufficient and bioreactor content was acidified, therefore, microbial fermentation slowed down considerably. The results thus do not reflect the optimal NAXUS fermentation by bacteria. However, we can still use the results for
verifying the butyrate sensor as nice difference in butyrate production were observed due to this
technical error. Indeed, after several cycles of addition of small intestinal content (at pH 6.5-7.0), the
pH again increased so that microbial activity and butyrate production suddenly increased again. It
would be interesting to confirm that the butyrate sensor can pick up these drastic changes in butyrate
levels.

\section{Discussion and Conclusion}
\label{sec:discussion}

The design of the sensor still has to be improved so that it would be possible to immediately connect
it to the colon vessel. We suggest an M20 connection that is able to screw on top of our lid (instead of
the red blind stopper). Attachment of the sensors to the PCB probably would then probably be better
through wires instead of a direct coupling. This avoids big device having to be coupled on the vessel
itself.
Despite these technical advances that still have to be made, the current experiment already showed
the potential usefulness of online metabolite detection through a biosensor. If the SCFA results could
have been read online than we could have been able to respond much faster to the human mistake
that was made (i.e. assigning the same pH probe to two different pH controllers) avoiding the huge
butyrate drop as observed in PC2. This data will still be very useful since a lot of different SCFA
concentrations and proportions can be linked with the readings of the biosensor. Analysis of these data
will also tell us if it would be possible to differentiate the different SCFA groups (AA, PA and BA). Finally, other types of sensors \cite{10} and/or learning algorithms  \cite{9} may also be considered.

\bibliographystyle{unsrt}  
\bibliography{references}  






\end{document}